\documentclass[a4paper]{article}

\usepackage{INTERSPEECH2019}

\usepackage[justification=centering]{caption} % to center the caption
\usepackage{multirow} % multirow in table
\usepackage{hyperref}

\newcommand{\x} {\vec{x}}
\newcommand{\y} {\vec{y}}
\newcommand{\h} {\vec{h}}
\newcommand{\vecq} {\vec{q}}
\newcommand{\vecc} {\vec{c}}
\newcommand{\veca} {\vec{a}}
\newcommand{\DTTS} {\vec{D}_{\text{TTS}}}
\newcommand{\DVC} {\vec{D}_{\text{VC}}}
\newcommand{\Enc} {\text{Enc}}
\newcommand{\Dec} {\text{Dec}}

\newcommand\blfootnote[1]{%
  \begingroup
  \renewcommand\thefootnote{}\footnote{#1}%
  \addtocounter{footnote}{-1}%
  \endgroup
}

\title{Voice Transformer Network: Sequence-to-Sequence Voice Conversion Using Transformer with Text-to-Speech Pretraining}
\name{\begin{tabular}{c}
	Wen-Chin Huang$^1$, Tomoki Hayashi$^1$, Yi-Chiao Wu$^1$, Hirokazu Kameoka$^2$, Tomoki Toda$^1$	
	\end{tabular}
}
\address{
  $^1$Nagoya University, Japan 
  $^2$NTT Communication Science Laboratories, Japan}
\email{wen.chinhuang@g.sp.m.is.nagoya-u.ac.jp}

\begin{document}

\maketitle
\begin{abstract}
  We introduce a novel sequence-to-sequence (seq2seq) voice conversion (VC) model based on the Transformer architecture with text-to-speech (TTS) pretraining. Seq2seq VC models are attractive owing to their ability to convert prosody. While seq2seq models based on recurrent neural networks (RNNs) and convolutional neural networks (CNNs) have been successfully applied to VC, the use of the Transformer network, which has shown promising results in various speech processing tasks, has not yet been investigated. Nonetheless, their data-hungry property and the mispronunciation of converted speech make seq2seq models far from practical. To this end, we propose a simple yet effective pretraining technique to transfer knowledge from learned TTS models, which benefit from large-scale, easily accessible TTS corpora. VC models initialized with such pretrained model parameters are able to generate effective hidden representations for high-fidelity, highly intelligible converted speech. Experimental results show that such a pretraining scheme can facilitate data-efficient training and outperform an RNN-based seq2seq VC model in terms of intelligibility, naturalness, and similarity.\blfootnote{Preprint. Work in progress.}
\end{abstract}
\noindent\textbf{Index Terms}: Voice Conversion, Sequence-to-Sequence Learning, Transformer, pretraining

\section{Introduction}
\label{sec:intro}

Voice conversion (VC) aims to convert the speech from a source to that of a target without changing the linguistic content \cite{VC}. Conventional VC systems follow an \textit{analysis\textemdash conversion \textemdash synthesis} paradigm \cite{GMM-VC}. First, a high quality vocoder such as WORLD \cite{WORLD} or STRAIGHT \cite{STRAIGHT} is utilized to extract different acoustic features, such as spectral features and fundamental frequency (F0). These features are converted separately, and a waveform synthesizer finally generates the converted waveform using the converted features. Past VC studies have focused on the conversion of spectral features while only applying a simple linear transformation to F0. In addition, the conversion is usually performed frame-by-frame, i.e, the converted speech and the source speech are always of the same length. To summarize, the conversion of prosody, including F0 and duration, is overly simplified in the current VC literature.

This is where sequence-to-sequence (seq2seq) models \cite{S2S} can play a role. Modern seq2seq models, often equipped with an attention mechanism \cite{S2S-NMT-Bah, S2S-NMT-Luong} to implicitly learn the alignment between the source and output sequences, can generate outputs of various lengths. This ability makes the seq2seq model a natural choice to convert duration in VC. In addition, the F0 contour can also be converted by considering F0 explicitly (e.g, forming the input feature sequence by concatenating the spectral and F0 sequences) \cite{S2S-VC, ATT-S2S-VC, CONV-S2S-VC} or implicitly (e.g, using mel spectrograms as the input feature) \cite{S2S-iFLYTEK-VC, S2S-Text-VC, S2S-NP-VC, S2S-FAC-VC, S2S-Hier-VC, S2S-parrotron-VC}. Seq2seq VC can further be applied to accent conversion \cite{S2S-FAC-VC}, where the conversion of prosody plays an important role.

Existing seq2seq VC models are based on either recurrent neural networks (RNNs) \cite{S2S-VC, ATT-S2S-VC, S2S-iFLYTEK-VC, S2S-Text-VC, S2S-NP-VC, S2S-FAC-VC, S2S-Hier-VC, S2S-parrotron-VC} or convolutional neural networks (CNNs) \cite{CONV-S2S-VC}. In recent years, the Transformer architecture \cite{transformer} has been shown to perform efficiently \cite{RNNvsTransformer} in various speech processing tasks such as automatic speech recognition (ASR) \cite{transformer-asr}, speech translation (ST) \cite{transformer-st, transformer-st-enhance}, and text-to-speech (TTS) \cite{transformer-tts}. On the basis of attention mechanism solely, the Transformer enables parallel training by avoiding the use of recurrent layers, and provides a receptive field that spans the entire input by using multi-head self-attention rather than convolutional layers. Nonetheless, the above-mentioned speech applications that have successfully utilized the Transformer architecture all attempted to find a mapping between text and acoustic feature sequences. VC, in contrast, attempts to map between acoustic frames, whose high time resolution introduces challenges regarding computational memory cost and accurate attention learning.
%It is expected that using a Transformer based structure in VC can boost the performance.

Despite the promising results, seq2seq VC models suffer from two major problems. First, seq2seq models usually require a large amount of training data, although a large-scale parallel corpus, i.e, pairs of speech samples with identical linguistic contents uttered by both source and target speakers, is impractical to collect. Second, as pointed out in \cite{S2S-Text-VC}, the converted speech often suffers from  mispronunciations and other instability problems such as phonemes and skipped phonemes. Several techniques have been proposed to address these issues. In \cite{S2S-iFLYTEK-VC} a pretrained ASR module was used to extract phonetic posteriorgrams (PPGs) as an extra clue, whereas PPGs were solely used as the input in \cite{S2S-FAC-VC}. The use of context preservation loss and guided attention loss \cite{DC-GA-TTS} to stabilize training has also been proposed \cite{ATT-S2S-VC, CONV-S2S-VC}. Multitask learning and data augmentation were incorporated in \cite{S2S-Text-VC} using additional text labels to improve data efficiency, and linguistic and speaker representations were disentangled in \cite{S2S-NP-VC} to enable nonparallel training, thus removing the need for a parallel corpus. In \cite{S2S-parrotron-VC} a large hand-transcribed corpus was used to generate artificial training data from a TTS model for a many-to-one (normalization) VC model, where multitask learning was also used.

One popular means of dealing with the problem of limited training data is transfer leaning, where knowledge from massive, out-of-domain data is utilized to aid learning in the target domain. Recently, TTS systems, especially neural seq2seq models, have enjoyed great success owing to the vast large-scale corpus contributed by the community. We argue that lying at the core of these TTS models is the ability to generate effective intermediate representations, which facilitates correct attention learning that bridges the encoder and the decoder. Transfer learning from TTS has been successfully applied to tasks such as speaker adaptation \cite{adaptation-cloning, adaptation-verification, adaptation-WN, espnet-tts}. In \cite{adaptation-vc} the first attempt to apply this technique to VC was made by bootstrapping a nonparallel VC system from a pretrained speaker-adaptive TTS model.

\begin{figure}[t]
  \centering
  \includegraphics[width=0.5\textwidth]{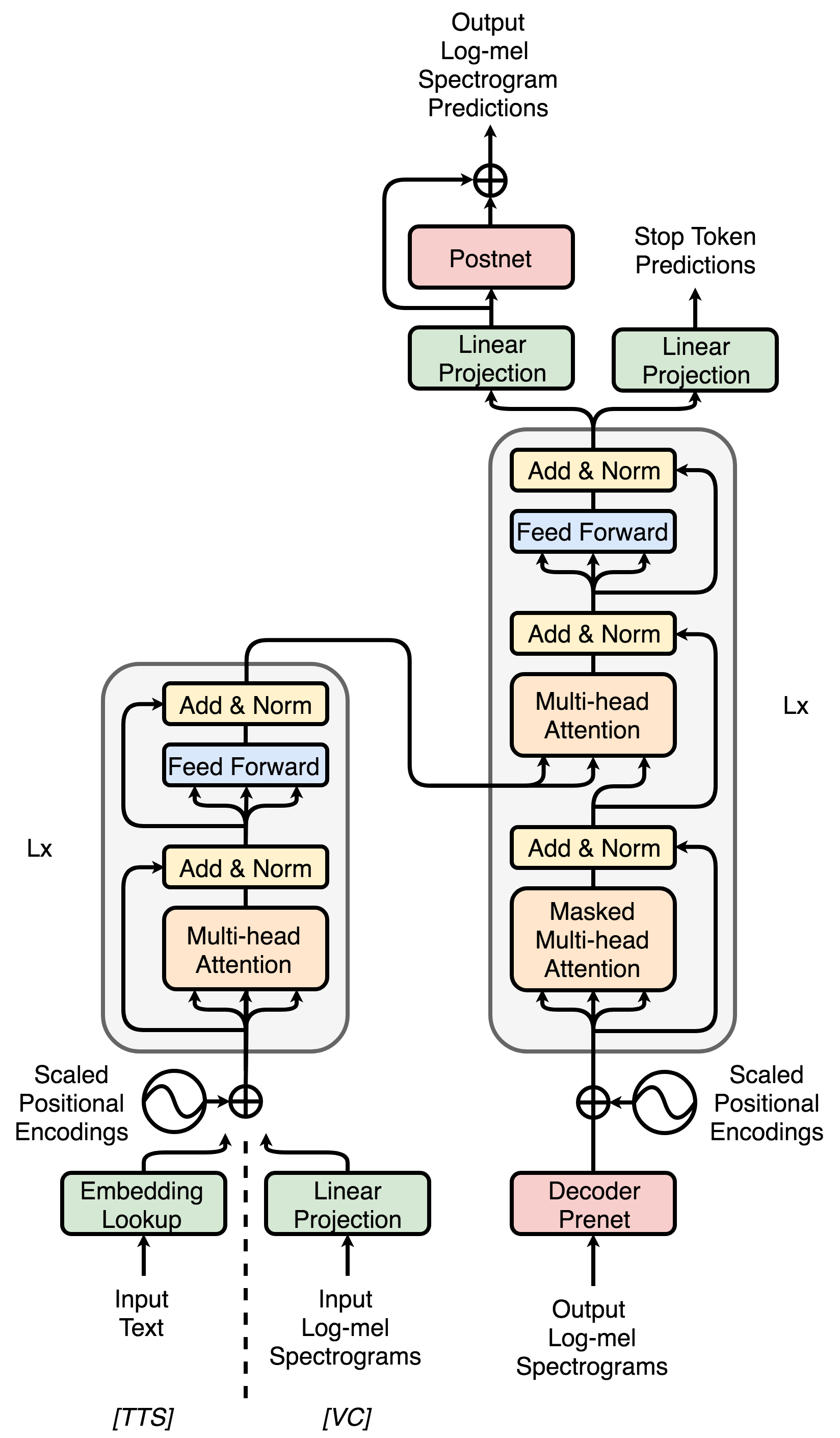}
  \centering
  \captionof{figure}{Model architecture of Transformer-TTS and VC.}
  \label{fig:transformer-ss}
\end{figure}

In this work, we propose a novel yet simple pretraining technique to transfer knowledge from learned TTS models. To transfer the core ability, i.e, the generation and utilization of fine representations, knowledge from both the encoder and the decoder is needed. Thus, we pretrain them in separate steps: first, the decoder is pretrained by using a large-scale TTS corpus to train a conventional TTS model. The TTS training ensures a well-trained decoder that can generate high-quality speech with the correct hidden representations. As the encoder must be pretrained to encode input speech into hidden representations that can be recognized by the decoder, we train the encoder in an autoencoder style with the pretrained decoder fixed. This is carried out using a simple reconstruction loss. We demonstrate that the VC model initialized with the above pretrained model parameters can generate high-quality, highly intelligible speech even with very limited training data.

Our contributions in this work are as follows:
\begin{itemize}
  \item We apply the Transformer network to VC. To our knowledge, this is the first work to investigate this combination.
  \item We propose a TTS pretraining technique for VC. The pretraining process provides a prior for fast, sample-efficient VC model learning, thus reducing the data size requirement and training time. In this work, we verify the effectiveness of this scheme by transferring knowledge from Transformer-based TTS models to a Transformer-based VC model.
\end{itemize}
%Our proposed TTS pretraining technique is most similar to that introduced in \cite{S2S-NP-VC}, where they also utilized text data and employed a two stage pretraining+finetuning strategy. Our method differs in that we do not require carefully designed network modules and loss functions, but rather make minimal changes to existing TTS models. Since our method does not rely on an specific TTS structure, it is expected that as more powerful TTS models are developed, the performance of our method also improves. 
%Our TTS pretraining method is most similar to that introduced in \cite{S2S-Hier-VC}, where they proposed autoencoder-style pretraining using the speech data in large scale TTS corpus. While the effectiveness of their approach is unknown due to the lack of objective or subjective evaluation results, a fundamental difference is that we additionally utilized the text data such that fine-grained hidden representations is ensured to be able to be generated and recognized.

\section{Background}
\subsection{Sequence-to-sequence speech systhesis}
\label{ssec:s2s-ss}

Seq2seq models are used to find a mapping between a source feature sequence $\x_{1:n}=(\x_1, \cdots, \x_n)$ and a target feature sequence $\y_{1:m}=(\y_1, \cdots, \y_m)$ which do not necessarily have to be of the same length, i.e, $n \neq m$. Most seq2seq models have an encoder\textemdash decoder structure \cite{S2S}, where advanced ones are equipped with an attention mechanism \cite{S2S-NMT-Bah, S2S-NMT-Luong}. First, an encoder ($\Enc$) maps $\x_{1:n}$ into a sequence of hidden representations $\h_{1:n}=(\h_1, \cdots, \h_n)$. The decoding of the output sequence is autoregressive, which means that the previously generated symbols are considered an additional input at each decoding time step. To decode an output feature $\y_t$, a weighted sum of $\h_{1:n}$ first forms a context vector $\vecc_t$, where the weight vector is represented by a calculated attention probability vector $\veca_t=(a^{(1)}_t, \cdots, a^{(n)}_t)$. Each attention probability $a^{(k)}_t$ can be thought of as the importance of the hidden representation $\h_k$ at the $t$th time step. Then the decoder ($\Dec$) uses the context vector $\vecc$ and the previously generated features $\y_{1:t-1}=(\y_1, \cdots, \y_{t-1})$ to decode $\y_t$. Note that both the calculation of the attention vector and the decoding process take the previous hidden state of the decoder $\vecq_{t-1}$ as the input. The above-mentioned procedure can be formulated as follows:
\begin{align}
	\h_{1:n} &= \Enc(\x_{1:n}),\\
	\veca_t &= \text{attention}(\vecq_{t-1}, \h_{1:n}),\\
	\vecc_t &= \sum_{k=1}^{n} a^{(n)}_t \h_k,\\
	\y_t , \vecq_t &= \Dec(\y_{1:t-1}, \vecq_{t-1}, \vecc_t).
\end{align}
As pointed out in \cite{adaptation-vc, joint-tts-vc}, TTS and VC are similar since the output in both tasks is a sequence of acoustic features. In such seq2seq speech synthesis tasks, it is a common practice to employ a linear layer to further project the decoder output to the desired dimension. During training, the model is optimized via backpropagation using an L1 or L2 loss.

\subsection{Transformer-based text-to-speech synthesis}
\label{ssec:transformer-tts}

In this subsection we describe the Transformer-based TTS system proposed in \cite{transformer-tts}, which we will refer to as Transformer-TTS. Transformer-TTS is a combination of the Transformer \cite{transformer} architecture and the Tacotron 2 \cite{Taco2} TTS system.

We first briefly introduce the Transformer model \cite{transformer}. The Transformer relies solely on a so-called multi-head self-attention module that learns sequential dependences by jointly attending to information from different representation subspaces. The main body of Transformer-TTS resembles the original Transformer architecture, which, as in any conventional seq2seq model, consists of an encoder stack and a decoder stack that are composed of $L$ encoder layers and $L$ decoder layers, respectively. An encoder layer contains a multi-head self-attention sublayer followed by a positionwise fully connected feedforward network. A decoder layer, in addition to the two sub-layers in the encoder layer, contains a third sub-layer, which performs multi-head attention over the output of the encoder stack. Each layer is equipped with residual connections and layer normalization. Finally, since no recurrent relation is employed, sinusoidal positional encoding \cite{convs2s} is added to the inputs of the encoder and decoder so that the model can be aware of information about the relative or absolute position of each element.

The model architecture of Transformer-TTS is depicted in Figure~\ref{fig:transformer-ss}. Since the Transformer architecture was originally designed for machine translation, several changes have been made to the architecture in \cite{transformer-tts} to make it compatible in the TTS task. First, as in Tacotron 2, prenets are added to the encoder and decoder sides. Since the text space and the acoustic feature space are different, the positional embeddings are employed with corresponding trainable weights to adapt to the scale of each space. In addition to the linear projection to predict the output acoustic feature, an extra linear layer is added to predict the stop token \cite{Taco2}. A weighted binary cross-entropy loss is used so that the model can learn when to stop decoding. As a common practice in recent TTS models, a five-layer CNN postnet predicts a residual to refine the final prediction.

In this work, our implementation is based on the open-source ESPnet-TTS \cite{espnet, espnet-tts}, where the encoder prenet is discarded and the guided attention loss is applied \cite{DC-GA-TTS} to partial heads in partial decoder layers \cite{RNNvsTransformer}.

\begin{figure}[t]
  \centering
  \includegraphics[width=0.48\textwidth]{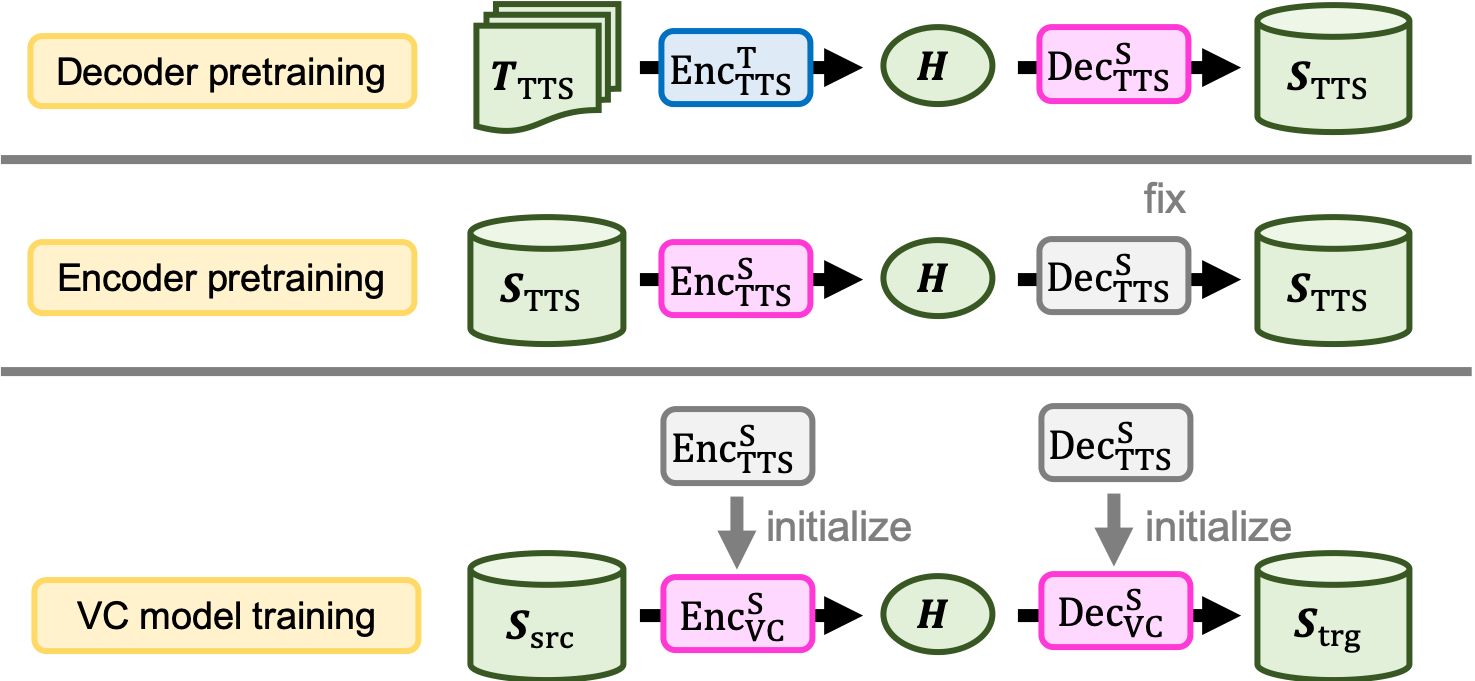}
  \centering
  \captionof{figure}{Illustration of proposed TTS pretraining technique for VC.}
  \label{fig:tts-pt}
\end{figure}

\section{Voice Transformer Network}
\label{sec:voice-transformer-network}

In this section we describe the combination of Transformer and seq2seq VC. Our proposed model, called the Voice Transformer Network (VTN), is largely based on Transformer-TTS introduced in Section~\ref{ssec:transformer-tts}. Our model consumes the source log-mel spectrogram and outputs the converted log-mel spectrogram. As pointed out in Section~\ref{ssec:s2s-ss}, TTS and VC respectively encode text and acoustic features to decode acoustic features. Therefore, we make a very simple modification to the TTS model, which is to replace the embedding lookup layer in the encoder with a linear projection layer, as shown in Figure~\ref{fig:transformer-ss}. Although more complicated networks can be employed, we found that this simple design is sufficient to generate satisfying results. The rest of the model architecture as well as the training process remains the same as that for Transformer-TTS.

%We utilized the ESPnet-TTS toolkit as a unified implementation and training framework, ensuring fair comparison and benchmarking.

An important trick we found to be useful here is to use a reduction factor in both the encoder and the decoder for accurate attention learning. In seq2seq TTS, since the time resolution of acoustic features is usually much larger than that of the text input, a reduction factor $r_d$ is commonly used on the decoder side \cite{Taco}, where multiple stacked frames are decoded at each time step. On the other hand, although the input and output of VC are both acoustic features, the high time resolution (about 100 frames per second) not only makes attention learning difficult but also increases the training memory footprint. While pyramid RNNs were used to reduce the time resolution in \cite{S2S-iFLYTEK-VC}, here we simply introduce an encoder reduction factor $r_e$, where adjacent frames are stacked to reduce the time axis. We found that this not only leads to better attention alignment but also reduces the training memory footprint by half and subsequently the number of required gradient accumulation steps \cite{espnet-tts}.

\begin{table*}[t]
	\centering
	\captionsetup{justification=centering}
	\caption{Validation-set objective evaluation results of adapted TTS, baseline (ATTS2S), and variants of the VTN trained on different sizes of data.}
	\centering
	\begin{tabular}{ c c c c c c c c c c c c c }
		\toprule
		 & & \multicolumn{2}{c}{pretraining} & \multicolumn{3}{c}{932 training utterances} & \multicolumn{3}{c}{250 training utterances} & \multicolumn{3}{c}{80 training utterances} \\
		\cmidrule(lr){3-4} \cmidrule(lr){5-7} \cmidrule(lr){8-10} \cmidrule(lr){11-13}
		Speaker & \multicolumn{1}{c}{Description} & Encoder & Decoder & MCD & CER & WER & MCD & CER & WER & MCD & CER & WER \\
		\midrule
		\multirow{4}{*}[-1pt]{clb-slt} & \multirow{3}{*}{VTN} & - & - & 7.16 & 15.1 & 23.2 & - & - & - & - & - & - \\
		& & - & \checkmark & 6.84 & 9.9 & 18.1 & 8.20 & 52.3 & 71.7 & - &  - & - \\
		& & \checkmark & \checkmark & \textbf{6.44} & \textbf{2.4} & \textbf{6.3} & \textbf{6.82} & \textbf{2.9} & \textbf{7.0} & \textbf{7.15} & \textbf{10.6} & \textbf{16.1} \\
		\cmidrule(lr){2-13} 
		& ATTS2S & - & - & 7.18 & 7.5 & 14.1 & 7.83 & 19.9 & 30.4 & 8.46 & 40.4 & 58.5 \\
		\midrule
		\multirow{4}{*}[-1pt]{bdl-slt} & \multirow{3}{*}{VTN} & - & - & 7.32 & 20.9 & 32.6 & - & - & - & - & - & - \\
		& & - & \checkmark & 7.07 & 20.3 & 31.7 & 8.52 & 58.9 & 80.7 & - &  - & - \\
		& & \checkmark & \checkmark & \textbf{6.56} & \textbf{6.1} & \textbf{12.0} & \textbf{7.14} & \textbf{9.6} & \textbf{16.2} & \textbf{7.34} & \textbf{15.0} & \textbf{23.7} \\
		\cmidrule(lr){2-13} 
		& ATTS2S & - & - & 7.36 & 13.0 & 22.7 & 8.01 & 28.4 & 43.9 & 8.76 & 51.2 & 72.7 \\
		\midrule
		slt & TTS adaptation & - & - & - & 3.3 & 6.2 & - & 4.5 & 7.2 & - & 4.8 & 7.8 \\
		\bottomrule
	\end{tabular}
	\label{tab:obj-eval}
\end{table*}

\section{Proposed training strategy with text-to-speech pretraining}
\label{tts-pt}

We present a text-to-speech pretraining technique that enables fast, sample-efficient training without introducing additional modification or loss to the original model structure or training loss. Assume that, in addition to a small, parallel VC dataset $\DVC=\{\vec{S}_{\text{src}}, \vec{S}_{\text{trg}}\}$, access to a large single-speaker TTS corpus $\DTTS=\{\vec{T}_{\text{TTS}}, \vec{S}_{\text{TTS}}\}$ is also available. $\vec{S}_{\text{src}}, \vec{S}_{\text{trg}}$ denote the source, target speech respectively, and $\vec{T}_{\text{TTS}}, \vec{S}_{\text{TTS}}$ denote the text and speech of the TTS speaker respectively. Our setup is highly flexible in that we do not require any of the speakers to be the same, nor any of the sentences between the VC and TTS corpus to be parallel. We employ a two-stage training procedure, where in the first stage we use $\DTTS$ to learn the initial parameters as a prior, and then use $\DVC$ to adapt to the VC model in the second stage. As argued in Section~\ref{sec:intro}, the ability to generate fine-grained hidden representations $\vec{H}$ is the key to a good VC model, so our goal is to find a set of prior model parameters to train the final encoder $\Enc^{\text{S}}_{\text{VC}}$ and decoder $\Dec^{\text{S}}_{\text{VC}}$. The overall procedure is depicted in Figure~\ref{fig:tts-pt}.

%Existing works on seq2seq VC models use 1000 training utterances (roughly 1 hour of speech data). This number, however, is far from sufficient for training the Voice Transformer Network, as we will show in Section~\ref{sec:exp}. One possible reason is that the Transformer contains more model parameters, making it more data-hungry. Although the techniques described in Section~\ref{sec:intro} can be applied to relieve this problem, we demonstrate that the training scheme introduced in this section, which we believe can benefit any seq2seq VC model but not restricted to the Voice Transformer Network, is extremely effective.

\subsection{Decoder pretraining}
\label{ssec:dpt}

The decoder pretraining is as simple as training a conventional TTS model using $\DTTS$. Since text itself contains pure linguistic information, the text encoder $\Enc^{\text{T}}_{\text{TTS}}$ here is ensured to learn to encode an effective hidden representation that can be consumed by the decoder $\Dec^{\text{S}}_{\text{TTS}}$. Furthermore, by leveraging the large-scale corpus, the decoder is expected to be more robust by capturing various speech features, such as articulation and prosody.

%\subsubsection{Decoder double pretraining}
%If text transcription of the target training utterances $T_{trg}$ is available, decoder double pretraining can be further conducted. As speaker adaptation in TTS has been shown to be successful in recent years \cite{adaptation-cloning, adaptation-verification, adaptation-WN, espnet-tts}, we perform adaptation using the $\{T_{trg}, M_{trg}\}$ dataset pair, shown in Figure~\ref{fig:tts-pt} (Step 2). 

\subsection{Encoder pretraining}
\label{ssec:ept}

A well pretrained encoder should be capable of encoding acoustic features into hidden representations that are recognizable by the pretrained decoder. With this goal in mind, we train an autoencoder whose decoder is the one pretrained in Section~\ref{ssec:dpt} and kept fixed during training. The desired pretrained encoder $\Enc^{\text{S}}_{\text{TTS}}$ can be obtained by minimizing the reconstruction loss of $\vec{S}_{\text{TTS}}$. As the decoder pretraining process described in Section~\ref{ssec:dpt} takes a hidden representation encoded from text as the input, fixing it in the encoder pretraining process guarantees the encoder to behave similarly to the text encoder $\Enc^{\text{T}}_{\text{TTS}}$, which is to extract fine-grained, linguistic-information-rich representations.

\subsection{VC model training}
\label{ssec:vc-training}

Finally, using $\DVC$, we train the desired VC models, with the encoder and decoder initialized with $\Enc^{\text{S}}_{\text{TTS}}$ and $\Dec^{\text{S}}_{\text{TTS}}$ pretrained in Section~\ref{ssec:ept} and Section~$\ref{ssec:dpt}$, respectively. The pretrained model parameters serve as a very good prior to adapt to the relatively scarce VC data, as we will show later. Also, compared with training from scratch, the model takes less than half the training time to converge with the pretraining scheme, enabling extremely efficient training.

%\begin{table}[t]
%	\centering
%	\captionsetup{justification=centering}
%	\caption{ASR objective evaluation results of the ground truth and the TTS models.}
%	\centering
%	\begin{tabular}{ l c c c c }
%		\toprule
%		& \multicolumn{2}{c}{validation} & \multicolumn{2}{c}{evaluation} \\
%		\cmidrule(lr){2-3} \cmidrule(lr){4-5}
%		\multicolumn{1}{c}{Description} & CER & WER & CER & WER \\
%		\midrule
%		Ground truth & 1.0 & 3.9 & 0.9 & 3.8 \\
%		TTS adaptation & 3.3 & 6.2 & 3.0 & 6.2 \\
%		\bottomrule
%	\end{tabular}
%	\label{tab:obj-eval-asr}
%\end{table}

%\begin{table}[t]
%	\centering
%	\captionsetup{justification=centering}
%	\caption{Evaluation set subjective evaluation results with 95\% confidence intervals of the ground truth, the TTS model, the baseline ATTS2S and the proposed model, VTN. The numbers in the parenthesis indicate the number of training utterances.}
%	\centering
%	\begin{tabular}{ l c c }
%		\toprule
%		Description & Naturalness & Similarity \\
%		\midrule
%		Ground truth & 4.65 $\pm$ 0.19 & - \\
%		TTS adaptation (932) & 3.80 $\pm$ 0.17 & - \\
%		\midrule
%		ATTS2S (932) & 2.67 $\pm$ 0.23 & 50\% $\pm$ 16\% \\
%		VTN (932) & 4.02 $\pm$ 0.17 & 83\% $\pm$ 9\% \\
%		VTN (80) & 3.75 $\pm$ 0.17 & 72\% $\pm$ 12\% \\
%		\bottomrule
%	\end{tabular}
%	\label{tab:sub-eval}
%\end{table}

\section{Experimental evaluation}
\label{sec:exp}

\subsection{Experimental settings}

We conducted our experiments on the CMU ARCTIC database \cite{CMU-Arctic}, which contains parallel recordings of professional US English speakers sampled at 16 kHz. One female (\textit{slt}) was chosen as the target speaker and one male (\textit{bdl}) and one female (\textit{clb}) were chosen as sources. We selected 100 utterances each for validation and evaluation, and the other 932 utterances were used as training data. For the TTS corpus, we chose a US female English speaker (\textit{judy bieber}) from the M-AILABS speech dataset \cite{M-AILABS} to train a single-speaker Transformer-TTS model. With the sampling rate also at 16 kHz, the training set contained 15,200 utterances, which were roughly 32 hours long.

The entire implementation was carried out on the open-source ESPnet toolkit \cite{espnet-tts, espnet}, including feature extraction, training and benchmarking. We extracted 80-dimensional mel spectrograms with 1024 FFT points and a 256 point frame shift. The base settings for the TTS model and training follow the \textit{Transformer.v1} configuration in \cite{espnet-tts}, and we made minimal modifications to it for VC. The reduction factors $r_e, r_d$ are both 2 in all VC models. For the waveform synthesis module, we used Parallel WaveGAN (PWG) \cite{parallel-wavegan}, which is a non-autoregressive variant of the WaveNet vocoder \cite{wavenet, sd-wnv} and enables parallel, faster than real-time waveform generation\footnote{We followed the open-source implementation at \url{https://github.com/kan-bayashi/ParallelWaveGAN}}. Since speaker-dependent neural vocoders outperform speaker-independent ones \cite{si-wnv}, we trained a speaker-dependent PWG by conditioning on natural mel spectrograms using the full training data of \textit{slt}. Our goal here is to demonstrate the effectiveness of our proposed method, so we did not train separate PWGs for different training sizes of the TTS/VC model used, although target speaker adaptation with limited data in VC can be used \cite{VC-WNV-adapt}.

We carried out two types of objective evaluations between the converted speech and the ground truth: the mel cepstrum distortion (MCD), a commonly used measure of spectral distortion in VC, and the character error rate (CER) as well as the word error rate (WER), which estimate the intelligibility of the converted speech. We used the WORLD vocoder \cite{WORLD} to extract 24-dimensional mel cepstrum coefficients with a 5 ms frame shift, and calculated the distortion of nonsilent, time-aligned frame pairs. The ASR engine is based on the Transformer architecture \cite{transformer-asr} and is trained using the LibriSpeech dataset \cite{librispeech}. The CER and WER for the ground-truth evaluation set of \textit{slt} were 0.9\% and 3.8\%, respectively. We also reported the ASR results of the TTS model adapted on different sizes of \textit{slt} training data in Table~\ref{tab:obj-eval}, which can be regarded as upper bounds.

\begin{table*}[t]
	\centering
	\captionsetup{justification=centering}
	\caption{Evaluation-set subjective evaluation results with 95\% confidence intervals for the ground truth, the TTS model, the baseline ATTS2S, and the proposed model, VTN. The numbers in the parentheses indicate the numbers of training utterances.}
	\centering
	\begin{tabular}{ l c c c c c c }
		\toprule
		 & \multicolumn{3}{c}{Naturalness} & \multicolumn{3}{c}{Similarity} \\
		\cmidrule(lr){2-4} \cmidrule(lr){5-7}
		Description & clb-slt & bdl-slt & Avg. & clb-slt & bdl-slt & Avg.  \\
		%Description & Nat. & Sim. & Nat. & Sim. & Nat. & Sim. \\
		\midrule
		Ground truth & \multicolumn{2}{c}{-} & 4.65 $\pm$ 0.19 & \multicolumn{3}{c}{-} \\
		TTS adaptation (932) & \multicolumn{2}{c}{-} & 3.80 $\pm$ 0.17 & \multicolumn{3}{c}{-} \\
		\midrule
		ATTS2S (932) & 2.83 $\pm$ 0.23 & 2.51 $\pm$ 0.27 & 2.67 $\pm$ 0.23 & 52\% $\pm$ 20\% & 48\% $\pm$ 13\% & 50\% $\pm$ 16\% \\
		VTN (932) & 3.94 $\pm$ 0.17 & 4.10 $\pm$ 0.17 & \textbf{4.02 $\pm$ 0.17} & 84\% $\pm$ 11\% & 82\% $\pm$ 10\% & \textbf{83\% $\pm$ 9\%} \\
		VTN (80) & 3.77 $\pm$ 0.17 & 3.72 $\pm$ 0.24 & \textbf{3.75 $\pm$ 0.17} & 65\% $\pm$ 17\% & 79\% $\pm$ 12\% & \textbf{72\% $\pm$ 12}\% \\
		\bottomrule
	\end{tabular}
	\label{tab:sub-eval}
\end{table*}

\subsection{Effectiveness of TTS pretraining}
\label{ssec:effectiveness-tts-pt}

To evaluate the importance and the effectiveness of each pretraining scheme we proposed, we conducted a systematic comparison between different training processes and different sizes of training data. The objective results are in Table~\ref{tab:obj-eval}. First, when the network was trained from scratch without any pretraining, the performance was not satisfactory even with the full training set. With decoder pretraining, a performance boost in MCD was obtained, whereas the ASR results were similar. Nonetheless, as we reduced the training size, the performance dropped dramatically, a similar trend to that reported in \cite{S2S-NP-VC}. Finally, by incorporating encoder pretraining, the model exhibited a significant improvement in all objective measures, where the effectiveness was robust against the reduction in the size of training data. Note that in the \textit{clb-slt} conversion pair, our proposed method showed the potential to achieve extremely impressive ASR results comparable to the TTS upper bound.

\subsection{Comparison with baseline method}

Next, we compared our VTN model with an RNN-based seq2seq VC model called ATTS2S \cite{ATT-S2S-VC}. This model is based on the Tacotron model \cite{Taco} with the help of context preservation loss and guided attention loss to stabilize training and maintain linguistic consistency after conversion. We followed the configurations in \cite{ATT-S2S-VC} but used mel spectrograms instead of WORLD features.

The objective evaluation results of the baseline are reported in Table~\ref{tab:obj-eval}. For the different sizes of training data, our system not only consistently outperformed the baseline method but also remained robust, whereas the performance of the baseline method dropped dramatically as the size of training data was reduced. This proves that our proposed method can improve data efficiency as well as pronunciation. We also observed that when trained from scratch, our VTN model had a similar MCD and inferior ASR performance compared with the baseline. As the ATTS2S employed an extra mechanism to stabilize training, this result may indicate the superiority of using the Transformer architecture over RNNs. We leave rigorous investigation for future work.

Systemwise subjective tests on naturalness and conversion similarity were also conducted to evaluate the perceptual performance\footnote{The audio samples can be found at \url{https://unilight.github.io/Publication-Demos/publications/transformer-vc/}}. For naturalness, participants were asked to evaluate the naturalness of the speech by the mean opinion score (MOS) test on a five-point scale. For conversion similarity, each listener was presented a natural speech of the target speaker and a converted speech, and asked to judge whether they were produced by the same speaker with the confidence of the decision, i.e., sure or not sure. Ten non-native English speakers were recruited. 

Table~\ref{tab:sub-eval} shows the subjective results on the evaluation set. First, with the full training set, our proposed VTN model significantly outperformed the baseline ATTS2S by over one point for naturalness and 30\% for similarity. Moreover, when trained with 80 utterances, our proposed method showed only a slight drop in performance, and was still superior to the baseline method. This result justifies the effectiveness of our method and also showed that the pretraining technique can greatly increase data efficiency without severe performance degradation.

Finally, one interesting finding is that the VTN trained with the full training set also outperformed the adapted TTS model, while the VTN with limited data exhibited comparable performance. Considering that the TTS models in fact obtained good ASR results, we suspect that the VC-generated speech could benefit from encoding the prosody information from the source speech. In contrast, the lack of prosodic clues in the linguistic input in TTS reduced the naturalness of the generated speech.

\section{Conclusion}

In this work, we successfully applied the Transformer structure to seq2seq VC. Also, to address the problems of data efficiency and mispronunciation in seq2seq VC, we proposed the transfer of knowledge from easily accessible, large-scale TTS corpora by initializing the VC models with pretrained TTS models. A two-stage training strategy that pretrains the decoder and the encoder subsequently ensures that fine-grained intermediate representations are generated and fully utilized. Objective and subjective evaluations showed that our pretraining scheme can greatly improve speech intelligibility, and it significantly outperformed an RNN-based seq2seq VC baseline. Even with limited training data, our system can be successfully trained without significant performance degradation. In the future, we plan to more systematically examine the effectiveness of the Transformer architecture compared with RNN-based models. Extension of our pretraining methods to more flexible training conditions, such as nonparallel training \cite{S2S-NP-VC, adaptation-vc}, is also an important future task.

\section{Acknowledgements}

This work was supported in part by JST PRESTO Grant Number JPMJPR1657 and JST CREST Grant Number JPMJCR19A3, Japan.

\bibliographystyle{IEEEtran}

\bibliography{ref}
\end{document}